\documentclass[useAMS,usenatbib]{mn2e}
\usepackage{amsmath}
\usepackage{float}
\usepackage{color}
\usepackage{listings}
\usepackage{graphicx}
\usepackage{natbib}
\usepackage{multirow}
\usepackage{calligra}
\usepackage[T1]{fontenc}

\title{On the AGN radio luminosity distribution and the black hole fundamental plane}
\author[A. Bonchi et al.]
{A. Bonchi,$^1$ F. La Franca,$^{1}$\thanks{E-mail: lafranca@fis.uniroma3.it} G. Melini,$^1$ A. Bongiorno,$^2$ \& F. Fiore$^2$
\\
$^1$ Dipartimento di Fisica, Universit\`a Roma Tre,
Via della Vasca Navale 84, 00146, Roma, Italy\\
$^2$ INAF Osservatorio Astronomico di Roma, Via Frascati 33, Monte Porzio Catone, 00044, Italy
}

\begin{document}

\input{Step.sty}
\date{\today}
\maketitle
\label{firstpage}

\begin{abstract}
We have studied the dependence of the AGN nuclear  radio (1.4 GHz) luminosity on both the AGN 2-10 keV X-ray
and the host-galaxy K-band luminosity. A complete sample of 1268
X-ray selected AGN (both type 1 and type 2) has been used, which is the largest catalogue
of AGN belonging to statistically well defined samples where radio, X and K band information exists.
 At variance with previous studies, radio upper limits have
been statistically taken into account using  a Bayesian Maximum Likelihood fitting method. It resulted that a good fit is obtained
assuming a plane in the 3D L$_R$-L$_X$-L$_K$ space, namely logL$_R$= $\xi_X$logL$_X$+ $\xi_K$logL$_K$ + $\xi_0$,
having a $\sim$1 dex wide (1$\sigma$) spread in radio luminosity. As already shown by \citet{lafranca10}, no evidence of bimodality in the
radio luminosity distribution was found and therefore any definition of radio loudness in  AGN is arbitrary. 
Using scaling relations between the BH
mass and the host galaxy K-band luminosity, we have also derived a new estimate
of the BH fundamental plane (in the L$_{5 GHz}$-L$_X$-M$_{BH}$ space). Our analysis shows
that previous measures of the BH fundamental plane are biased by $\sim$0.8 dex in favor of the most luminous
radio sources. Therefore, many  AGN studies, where the BH fundamental plane is used to investigate
how AGN regulate their radiative and mechanical luminosity as a function of the accretion rate, or many AGN/galaxy co-evolution models, where radio-feedback is computed using the AGN fundamental plane,   should revise their conclusions.

\end{abstract}
\begin{keywords}
galaxies: active - radio continuum: galaxies - X-rays: galaxies  - methods: statistical.
\end{keywords}

\section{Introduction}

In the last years, in many galaxy formation models, Active Galactic Nuclei (AGN) have been considered related to
mechanisms capable to switch off the star formation in the most massive galaxies, thus reproducing both the observed shape of the galaxy luminosity function and the red, early type, passive  evolving nature of local massive galaxies. It is expected that  the AGN and galaxy evolutions are closely connected to each other (the AGN/galaxy co-evolution) through feedback processes coupling both the star formation and the black hole (BH) accretion rate histories.
Some of these models assume that the AGN feedback into the host galaxy is due to the kinetic energy
released by the radio jets and is, therefore,  dependent on the AGN radio luminosity \citep{croton06, cattaneo06, marulli08}. 
It has been, indeed, already demonstrated that the conversion of the AGN radio luminosity function
into a kinetic luminosity function provides the adequate amount of energy \citep{best06,merloni08,shankar08,kording08,cattaneo09,smolcic09,lafranca10}.

In this context, in order to build up more realistic AGN/galaxy co-evolutionary models, it is very useful to measure
the dependence of the AGN core radio luminosity on other physical quantities, related to the AGN/galaxy
evolutionary status, such as the BH and galaxy star masses and their time derivatives: accretion and star formation rates. 
These quantities  can be, either directly or indirectly, measured. 

A good estimate of galaxy star masses
can be obtained by spectral energy distribution (SED) analyses in the optical and near infrared (NIR) domains
\citep[see e.g.][]{merloni10,pozzi12}, or, still satisfactorily,
 from NIR (e.g. K-band) luminosity measures, as the mass-to-light ratio 
in the K band has a 1$\sigma$ scatter of 0.1 dex 
   \citep{madau98,bell2003}. 
 
BH masses can be estimated using reverberation mapping techniques or measuring the width of broad emission lines observed in optical and NIR spectra \citep[the single epoch method; see e.g.][]{vest02}. Less direct estimates are obtained using the scaling relations observed between the BH mass and the bulge or spheroid mass, or between the BH mass and the bulge luminosity of the host galaxies 
\citep[eg.][]{dressler89,kormendy93,kormendy95,mag98}.
In this framework, even the total host galaxy K-band luminosity, if converted into the bulge luminosity, can  be used as a good proxy  of the BH mass \citep[see e.g.][]{fiore12}.

The accretion rate, $\dot{m}$, is related to the
AGN hard (>2 keV) X-ray luminosity, L$_X$, via the
knowledge of the X-ray bolometric correction, K$_X$, and the efficiency,
$\epsilon$, of conversion of mass accretion into radiation,
\begin{equation}
L_X = {L_{bol}\over K_X}= {\epsilon \dot{m}{c^2}\over{(1-\epsilon) K_X}},
\end{equation}
where L$_{bol}$ is the bolometric luminosity and typical values for
$\epsilon$ are about $0.1$ \citep{mar04, vf09}. 

The AGN radio luminosity distribution and its relationship to either the optical or X-ray luminosity has been
studied by many authors.  Some of these studies discussed the AGN radio luminosity in terms of a bimodal distribution where two separate populations of radio loud and radio quiet  objects exist \citep[e.g.][]{kel89,miller90}. Many other studies have alternatively measured a relationship between the
radio and X-ray luminosities  \citep[e.g.][]{brink00,tw03,panessa07,bianchi09b,singal11}.  
 However almost all these studies are based on incomplete samples due to the  lack of deep radio observations. AGN samples selected in other bands (typically optical or X-ray) are therefore not fully detected in the radio band. 

However, in order to properly study the AGN radio luminosity properties and their relation to X-ray and optical luminosities, it is necessary to use fully (or almost) radio detected and complete AGN samples and when needed, to properly take into account the radio upper limits. 
More recently, using deep radio observations,  it  has been shown that both the AGN radio/optical and the radio/X-ray luminosity ratios span continuously more than 5 decades \citep[][]{best05,lafranca10,singal11,balokovic12}, without evidences of a bi-modal distributions,  and it is, therefore, inaccurate to deal with the AGN radio properties in terms of two separate populations of radio loud and radio quiet objects.   \citet{lafranca10} used a sample of about 1600 hard X-ray (mostly 2-10 keV) selected AGN to measure (taking also into account the presence of
censored radio data) the
probability distribution function (PDF) of the ratio between the nuclear radio (1.4 GHz) and
the X-ray luminosity R$_X$= log[$\nu$ L$_{\nu}$(1.4 GHz)/L$_X$(2-10 keV)] \citep[see][for a discussion on 
the difference between the radio to optical and the radio to X-ray ratio distributions in AGN]{tw03}.  The
probability distribution function of R$_X$ was functionally fitted as
dependent on the X-ray luminosity and redshift, $P(R_X| L_X, z)$.
The measure of the probability distribution function of R$_X$ eventually
allowed to compute the AGN kinetic luminosity function and the kinetic energy
density.

In this paper, using the same sample and a similar method as used by \citet{lafranca10},  we describe the measure of the dependence of the AGN radio core luminosity, L$_R$, distribution, on both the
X-ray (2-10 keV) luminosity, L$_X$, and the host galaxy (AGN subtracted) K-band luminosity, L$_K$. This measurement, in the context of the AGN/galaxy coevolution scenario (see above discussion), is vey useful in order to relate the kinetic (radio) feedback to the accretion rate (L$_X$) and the galaxy assembled star mass (L$_K$). In order to accurately take into account the
presence of censored data in the radio band, an ad hoc  Bayesian Maximum Likelihood (ML) method and a three dimensional Kolmogorov Smirnov test have been developed.

Many authors have observed the existence of an analogous  relationship between the radio luminosity,
the X-ray luminosity, and black hole mass ($M$), the BH fundamental plane,  namely 
${\rm log} L_R = \xi_{RX} {\rm log} L_X + \xi_{RM} {\rm log} M + constant$ 
\citep[see][]{merl03,falcke04,gult09}. The measure of such a relation is very useful in
order to discriminate among several theoretical models of jet production in AGN, as it suggests that
BH regulate their radiative and mechanical luminosity in the same way at any given accretion rate scaled
to Eddington \citep{falcke95, heinz03,churazov05,laor08}.

 As  relations have been observed between the BH mass and the K-band bulge luminosity (see discussion above),  in section 6 we convert our measure of 
the relation in the log$L_R$-log$L_X$-log$L_K$ space, into a relation into the
log$L_R$-log$L_X$-log$M$ space, and discuss
how much important is to properly  take into account the presence of radio upper limits in order
to measure the BH fundamental plane.

Unless otherwise stated, all quoted errors are at the 68\% confidence
level. We assume H$_0$=70 Km s$^{-1}$ Mpc$^{-1}$, $\Omega_m$=0.3 and
$\Omega_\Lambda$=0.7.

\section{The data}

  In our analysis we used the same data-set used by  \citet{lafranca10}, where radio (1.4 GHz) observations  (either detections or upper limits) were collected for 1641 AGN  (both type 1, AGN1,  and type 2, AGN2, i.e. showing or not the broad line region in their optical spectra) belonging to complete  (i.e. with almost all redshift and N$_H$ measures available)
 hard X-ray ($>$2 keV;  mostly 2-10 keV) selected AGN samples, with unabsorbed 2-10 keV 
luminosities larger
 than 10$^{42}$ erg/s\footnote{Throughout this work we  assumed that all the X-ray sources having 2-10 keV
unabsorbed luminosities larger  than 10$^{42}$ erg/s are AGN. See e.g. \citet{ranalli03} for a study of the typical X-ray luminosities of star forming galaxies.}.
  
   As the goal was to use the radio luminosity  in order to
  estimate the kinetic luminosity, \citet{lafranca10}
measured a radio emission which was as much as possible
  causally linked (contemporary) to the observed X-ray activity
  (accretion). Radio fluxes were measured  in a region
  as close as possible to the AGN, therefore minimizing the
  contribution of objects like the radio lobes in FRII sources
  \citep{fanaroff74}.  For this reason they built up a large
data-set of X-ray selected AGN (where redshift and N$_H$ column
densities estimates were available) observed at 1.4 GHz with a
$\sim$1\arcsec\ typical spatial resolution (1\arcsec\
corresponds, at maximum, to about 8 Kpc at $z$$\sim$2).  The cross
correlation of the X-ray and radio catalogues was carried out inside a
region with 5\arcsec\ of radius (almost less than, or equal to, the size
of the central part of a galaxy like ours), following a maximum
likelihood algorithm as described by \citet{sutherland92}
and \citet{ciliegi03}.  The
off-sets between the X-ray and radio positions of the whole sample resulted to have a root mean square (rms) of 1.4\arcsec
\citep[similar to the typical values obtained in X-ray to optical cross-correlations; e.g.][]{cocchia07}.

In order to measure the  K-band galaxy luminosity,  the   1641 AGN from \citet{lafranca10} have been cross-correlated with already existing K-band photometric catalogues  as explained below.

	\subsection{The local Sample: SWIFT and Grossan}

At the lowest redshift we have used  a sample of 33 AGN belonging to the 22 month SWIFT catalogue  \citep{tueller08}. These AGNs have been
detected at high galactic latitude ($|b|>15\deg$) in the  14-195 keV band with fluxes brighter than 10$^{-11}$ erg s$^{-1}$
cm$^{-2}$; all objects have N$_H$ column density and optical spectroscopic classification available.  The radio luminosity  at 1.4 GHz has been derived using  the
Faint Image of the Radio Sky at Twenty cm (FIRST)  Very Large Array (VLA) survey \citep{becker95}. In the case of no radio detection, a 5$\sigma$ upper limit of 0.75 mJy was adopted. The correlation between catalogues was made through the likelihood ratio technique \citep{sutherland92,ciliegi03}.
$K_S$ magnitudes have been associated to the sources using the 2MASS (Two Micron All Sky Survey) catalogue which has a  $\sim$14.3 mag completeness limit magnitude \citep{skrutskie06}. The cross-correlation with the K-band data has been carried out by comparing the positions of the the 2MASS sources on the K-band image  of the AGN counterpart.

To enlarge the local sample, \citet{lafranca10} used the hard X-ray selected AGN catalogue detected by the HEAO-1 mission (2-10 keV fluxes brighter than $2 \times 10^{-11}$ erg s$^{-1}$ cm$^{-2}$) described by \citet{grossan92} and revised by \citet{brusadin03}.  As for the SWIFT sample these sources have been cross correlated with the  FIRST and 2MASS radio and K-band catalogues,  respectively.

In summary, the local sample contains 43 X-ray sources all having a K$_S$ band detection.

	\subsection{HBSS}

We have used the 32 AGN selected by the Hard Bright Sensitivity Survey of the \emph{XMM-Newton} satellite, carried out at 4.5-7.5 keV fluxes brighter than 7$\times10^{-14}$ erg s$^{-1}$ cm$^{-2}$ \citep{dellaceca08}.  K$_S$ detections and upper limits were obtained for 19 and 13 sources, respectively,  from the 2MASS 
catalogue. The cross-correlation was carried using the same technique used for the local sample.
Those sources  missing a K$_S$ detection  were eventually excluded from the analysis.
	
	\subsection{The ASCA surveys: AMSS and ALSS}
	
Two samples come from observations of the ASCA satellite:
the ASCA Medium Sensitivity Survey \citep[AMSS;][]{akiyama03}, which is composed of 43 AGN, and the ASCA Large Sky Survey \citep[LASS;][]{ueda99}, which is composed of 30 AGN. For both of these catalogues we used the  K$_S$  photometric measures by \citet{watanabe04}. Only one source (belonging to the  AMSS) misses the K$_S$  detection.

	\subsection{COSMOS}	
COSMOS is the biggest catalogue used in this work. The catalogue by \citet{cappelluti09} of XMM-Newton sources with 2-10 keV fluxes brighter than  $\sim$3$\times 10^{-15}$ erg s$^{-1}$ cm$^{-2}$ was used. K-band magnitudes for 648 out of 677 sources were taken from  \citet[][and private communication]{brusa10}. 

\begin{table}
\begin{center}
\begin{tabular}{l c c c c}\hline
Sample	& $N$	& $N_{K}$	&$N_{K-GL}$ & N$_{R}$\\
& (1) & (2) & (3) & (4) \\
		& \scriptsize{La Franca (2010)}	& \scriptsize{K detected} & \scriptsize{K glx lum} &\scriptsize{Radio det}\\
\hline
SWIFT	  & ~~33	& ~~33	& ~~21	&~19 \\
GROSSAN & ~~10	& ~~10 	& ~~~0	&~~0\\
HBSS	  & ~~32	& ~~19 	& ~~17	&~~4\\
ALSS	  & ~~30	& ~~30 	& ~~19	&~~4\\
AMSS	  & ~~43	&~~42	& ~~19	&~~4\\
COSMOS   & ~677	& ~648 	& ~575	&121\\
CLANS	  & ~139	& ~125 	& ~~91	&~52\\
ELAIS	  & ~421	& ~363 	& ~283	&~37\\
CDF-S	  & ~~94	& ~~86	& ~~85	&~12\\
CDF-N 	  & ~162	& ~161 	& ~158	&~40\\
\hline
 Total		&1641	& 1517 &1268	&293\\
\end{tabular}
\end{center}
\caption{Samples breakdown}
\label{tabsample}
\end{table}

	\subsection{CLANS}
K-band photometry for 125 out of 139 AGN belonging to the Chandra Lockman Area North Survey were obtained from \citet{trouille08}. The radio data comes \citet{owen08}.

	\subsection{ELAIS-S1}
In the field S1 of the European Large Area ISO Survey  (ELAIS-S1) we used the catalogue of X-ray sources detected by XMM-Newton by \citet{puccetti06} which reaches a 2-10 keV flux limit of 2$\times 10^{-15}$ erg s$^{-1}$ cm$^{-2}$.  Radio data were taken from \citet{middelberg08}, while spectroscopic and photometric identifications were taken from \citet{lafranca04,berta06, feruglio08} and \citet{sacchi09}. In \citet{feruglio08} K$_S$ detections for 363 objects out of the 421 sources used by \citet{lafranca10} are available.

	\subsection{Deep samples: CDF South and North}
The deepest X-ray catalogues  used in this work are those available in the \emph{Chandra Deep Field South} (CDFS) and \emph{North} (CDFN). These are sub-samples of the GOOD-S and GOOD-N multi-wavelength surveys with 2-10 keV flux limits of $2.6\times 10^{-16}$ erg s$^{-1}$ cm$^{-2}$ and $1.4\times 10^{-16}$  erg s$^{-1}$ cm$^{-2}$, respectively. 

In the CDFS we  used 94 sources from the catalogue of \citet{alexander1303} and identified by \citet{brusa10}. These sources were  correlated with radio data taken from \citet{miller08}. 
K$_S$ band photometry was obtained using the deepest multi-wavelength catalogue (FIREWORKS) provided by
\citet{wuyts08} that reaches K$_S\simeq$22.5 mag. The  K$_S$ band  catalogue was cross correlated with the optical catalogue using the  likelihood ratio technique. K$_S$ band counterparts for 86 out of 94 sources were found.

In the CDFN we used the X-ray catalogue from Alexander et al. (2003) with the identifications from \citet{trouille08}.
The sample consists of 162 extragalactic sources for which radio informations were obtained from \citet{biggs06}. \citet{trouille08}  provided K$_S$ band detections for all but one of the sources. 

In Table \ref{tabsample} the breakdown of all samples used is reported. In column 1) it is shown the original number, N,  of the sources 
contained in the samples used by \citet{lafranca10}, while in column 2)  it is shown the number of sources, N$_K$,
having a K-band detection.

\section{K-band AGN and galaxy luminosities}

The K-band absolute magnitudes, M$_K$, have been computed by applying an empirical K-correction.
 We used the formula
\begin{equation}
M_K = m_K+5-5{\rm log}d_l(z)+ 2.5(1+\alpha){\rm log}(1+z),
\end{equation}
where $d_l(z)$ is the luminosity distance and the K-correction is represented by the $\alpha$ parameter.
We used $\alpha=-0.86$ after a comparison of our data with the COSMOS catalogue where absolute K-band magnitudes have been accurately computed through 
SED fitting techniques \citep[][and references therein]{bongiorno12}.

\begin{figure}
\begin{center}

\includegraphics[width = 8 cm]{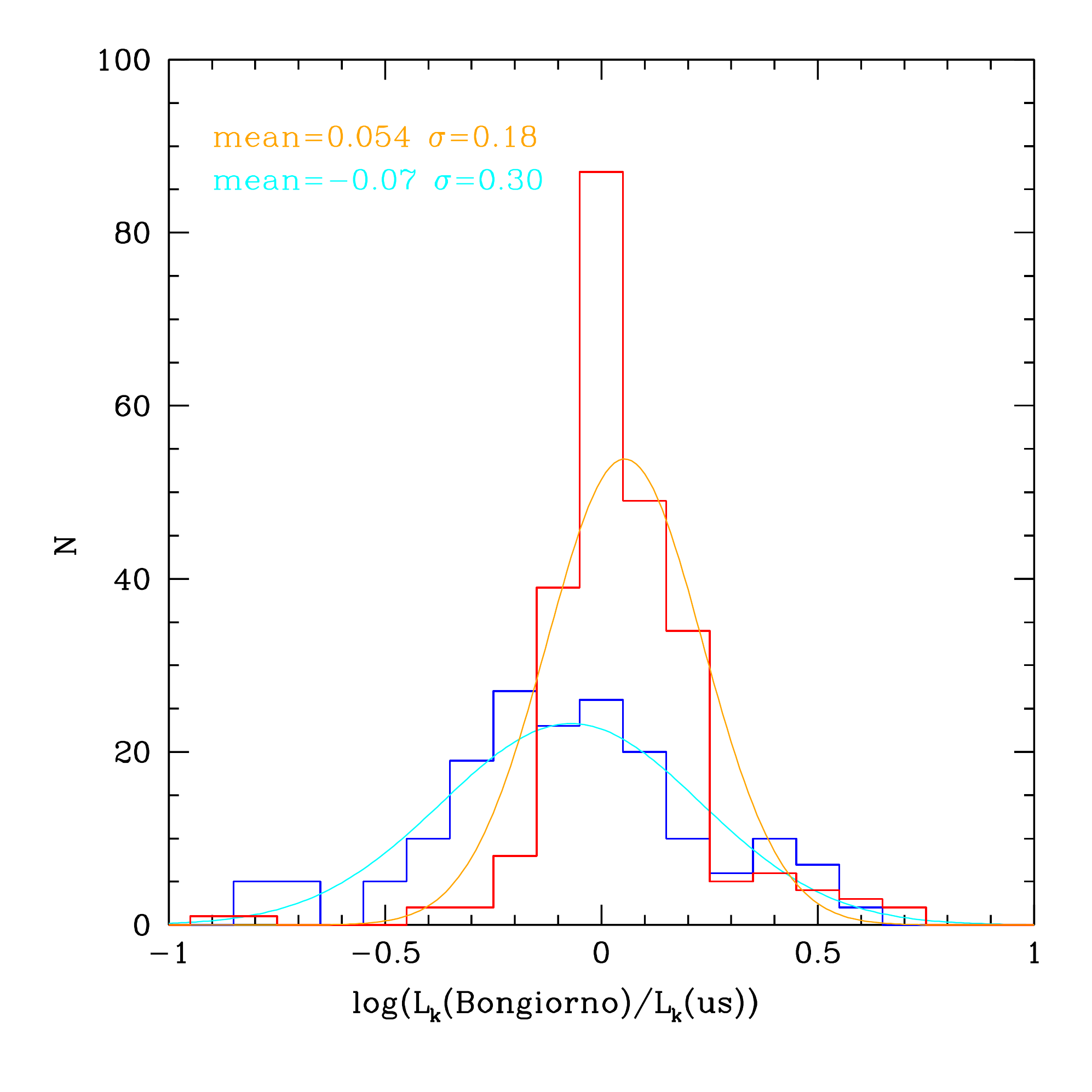}
\end{center}
\caption{Histogram of the logarithmic differences between our estimates of the K-band luminosities of the host galaxies (after the AGN component subtraction; see text) in the COSMOS sample and the
SED fitting measures from Bongiorno et al. (2012). AGN1 and AGN2 are shown by blue and red lines, respectively.
 }
\label{Fig_Angela}
\end{figure}

In order to estimate the host galaxy K-band luminosities (i.e. the stellar component) we subtracted the AGN contribution from the measured total luminosities. For this purpose we used the
nuclear (AGN only) infrared SEDs, normalized to the hard X-ray (2-10 keV) intrinsic luminosity and averaged within bins of absorbing N$_H$ as
published by \citet{silva04}. These AGN SEDs were obtained trough the interpolation, using 
updated models from \citet{granato94}, of the nuclear IR data of AGN taken from the \citet{maiolino95} sample.
The accuracy of our method has been tested by comparing our estimates with those obtained by \citet{merloni10} and
\citet{bongiorno12}  on the COSMOS catalogue using SED decomposition fitting  techniques.  As shown in Figure \ref{Fig_Angela}, our estimates, although less
accurate, are in good agreement with those obtained by \citet{merloni10} and \citet{bongiorno12}. The average difference results to be log$L_K$(COSMOS)-log$L_K$(us)=-0.07 (0.05) dex, with a 1$\sigma$ spread of 0.30 (0.18) dex for AGN1 (AGN2).
  
In some cases we obtained that the  expected AGN K-band luminosity was very close to (or even larger than) the total measured (AGN + host galaxy) luminosity. 
With the SED decomposition fitting  techniques used by \citet{merloni10} and \citet{bongiorno12}  no object
could result to have an AGN luminosity larger than the total one \citep[see also][]{pozzi07,pozzi12}. 
\citet{bongiorno12} conservatively decided that if the galaxy component were smaller than $10\%$  of the total one, only an upper limit, corresponding
to 10\% of  the total luminosity, could be assigned. Following this approach,  we decided to adopt 
a more conservative assumption, and  excluded from our analysis those 249 sources  where the galaxy component resulted to be smaller than $20\%$ of the total
one\footnote{ 
As described in section 5, our 3D ML fitting method is able to deal with upper limits on one physical quantity only (the
radio luminosity in our case). Therefore all sources where an upper limit on their K-band luminosity was available, were excluded from our analysis (this happened to all the 10 sources of the GROSSAN sample).}.

In Table \ref{tabsample} we report in column  3)  the number, $N_{K-GL}$, of sources where it was possible to estimate the galaxy K-band luminosity.

\begin{figure}
\begin{center}
\includegraphics[width = 6.5 cm,angle=-90]{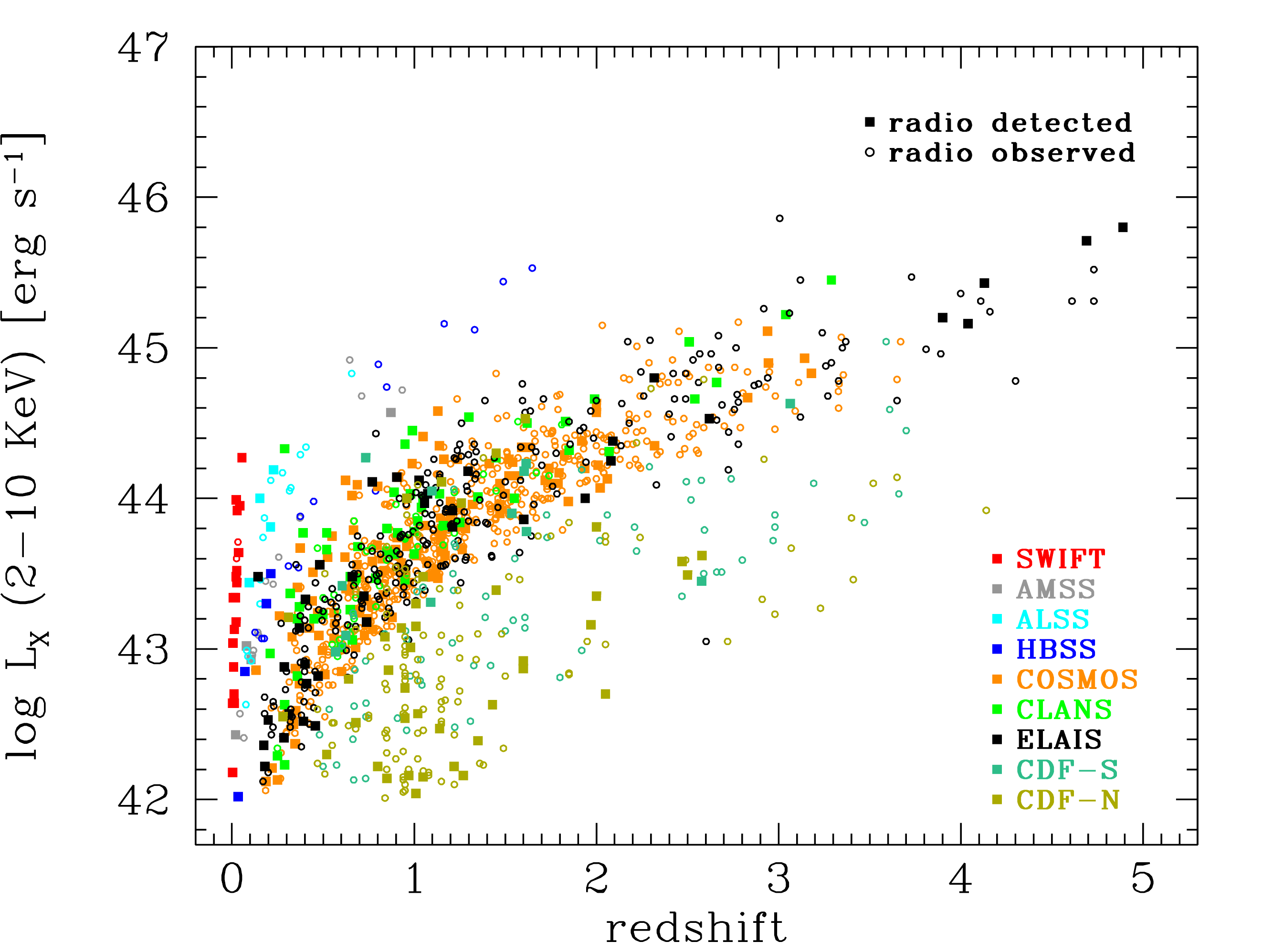}
\end{center}
\caption{2-10 keV de-absorbed luminosity, L$_X$, of the total sample as a function of redshift .}
\label{FigLz}
\end{figure}

\begin{figure}
\begin{center}
\includegraphics[width = 8.5 cm]{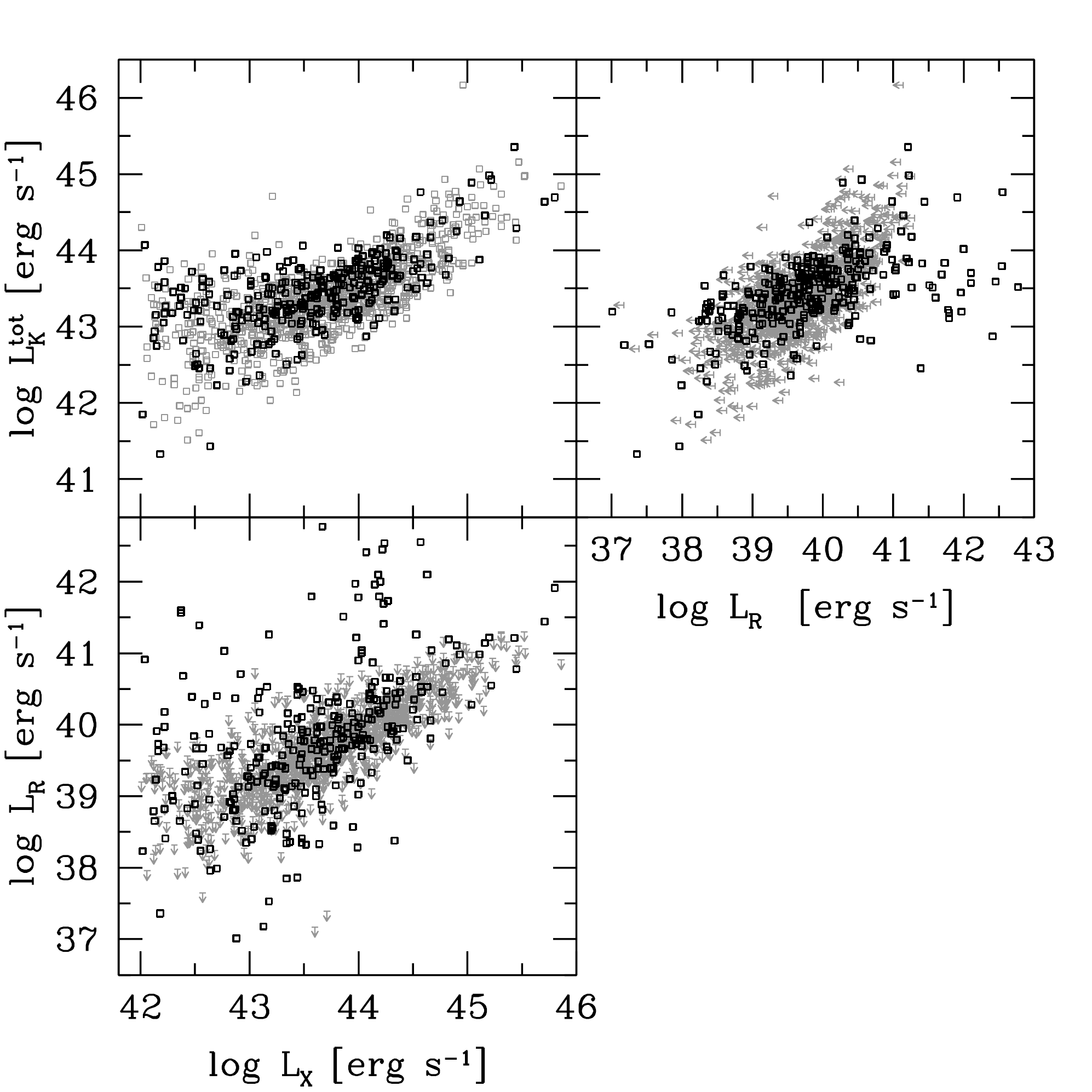}
\end{center}
\caption{Distribution of the sample onto the 2D L$_K$-L$_X$,
L$_K^{tot}$-L$_R$ and L$_X$-L$_R$  planes, where L$_K^{tot}$ is the total (galaxy + AGN) K-band luminosity. 
Squares represent the radio detected sources while
the sources with a radio upper limit are shown by either grey arrows or grey squares (upper left panel). }
\label{Fig_OrtNoSub}
\end{figure}

\begin{figure}
\begin{center}
\includegraphics[width = 8.5 cm]{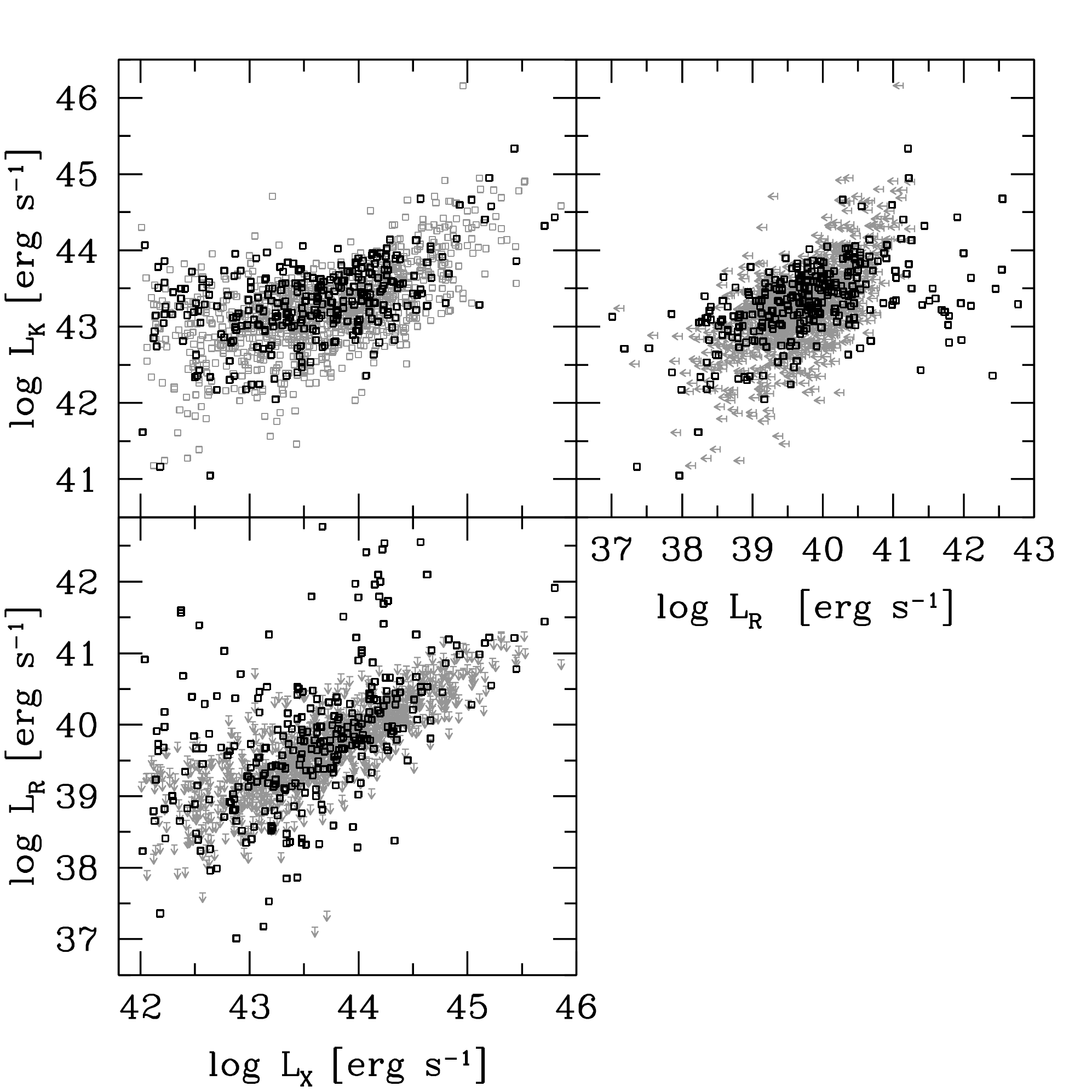}
\end{center}
\caption{As in Figure \ref{Fig_OrtNoSub} with L$_K$ representing the AGN subtracted K-band galaxy luminosity.}
\label{Fig_OrtSub}
\end{figure}

	\section{The whole sample}
	
In summary our data-set is composed by 9 X-ray selected AGN samples which contain a total of  1268 sources for which we were able to estimate
the K-band stellar component luminosities of the host galaxy, L$_K$.
For all these AGN,  column densities and de-absorbed 2-10 keV luminosities, L$_X$, measures are available. 
The radio data allowed to measure on 293 sources the ``nuclear'' 1.4 GHz luminosity, L$_R$, while for the remaining sources 5$\sigma$ radio upper limits (Table \ref{tabsample}, column 4) are available. In total, this is the largest catalogue
of AGN (both of type 1 and 2) belonging to statistically well defined samples where radio, X and K band information exist.
 The distribution, in the L$_X$-$z$ plane  of the whole AGN sample is shown in Figure \ref{FigLz},
while in Figure \ref{Fig_OrtNoSub}  and Figure  \ref{Fig_OrtSub} we show the 3D L$_K$-L$_X$-L$_R$  distribution, projected on to  the three 2D L$_K$-L$_X$,
L$_K$-L$_R$ and L$_X$-L$_R$  planes,  where L$_K$ is shown  before (L$_K^{tot}$), and after (L$_K$), the subtraction of the AGN component in the K-band, respectively.

\section{The plane fitting}

	\subsection{Maximum Likelihood fitting method}

We have used a maximum likelihood  fitting technique, with a Bayesian approach, in order to derive the probability distribution function of the AGN radio luminosity, L$_R$, as a function of  L$_X$ and the K band stellar component luminosity, L$_K$, $P(L_{R}  | L_{X}, L_{K})$. The maximum likelihood fitting method doesn't need to use  binning (as, e.g.,  it happens when using the $\chi^2$ fitting method), and therefore the results do not depend on the arrangement of the binning pattern.
In the usual maximum likelihood fitting of luminosity function distributions $\rho(z, L)$ of extragalactic sources 
\citep[eg.][]{marshall83} the best fit solution is  obtained by minimizing the quantity $S=-2ln\mathcal{L}$  \citep[where $\mathcal{L}$ is the likelihood,  and the $S$ function follows the $\chi^2$ statistic and therefore allows to estimates
the confidence interval of the best fit parameters;][]{lampton76}. The natural logarithm of the likelihood function is computed as follows:

\begin{equation}
ln\mathcal{L}= \sum_{i} ln(\rho(z_i,L_i) ) - 
    \int \rho(z,L) \Omega(z,L) {dV\over dz}dL dz,
    \label{ML1}
\end{equation}

where the sum is made over all the $i$ observed sources and $ \Omega(z,L) $ is the sky coverage as function of the 
luminosity $L$ and redshift $z$. The first term is proportional to the combined probability (of independent events)  of observing all the $i$ sources, each having redshift $z_i$ and luminosity $L_i$,
while the 
second term corresponds to the total number of expected sources in the sample and is, therefore, also used to constrain the normalization of the luminosity function. 

In our case we have devised a new $S$ function to be minimized which, following the
same statistic principles of the maximum likelihood method, is able to measure the conditional probability distribution function $P(L_{R}  | L_{X}, L_{K})$  (where ${\int} P(L_R  | L_{X}, L_{K}) dL_R = 1$)
of observing an $L_R$ radio luminosity in an object having $L_X$ and $L_K$ luminosities.
Our method has the advantage (in comparison to other three-dimensional fitting techniques) to be also able to take into account the occurrence of upper limits in one of the three dimensions.
Indeed, as already discussed, in all samples the radio observations are not deep enough do detect all the sources   
(see Table \ref{tabsample}) and therefore upper limits need to be considered in order to derive the true
AGN radio luminosity distribution \citep[see][for an analogous Bayesian approach on this topic]{plotkin12}.  For these reasons the  natural logarithm of the likelihood function
has been  computed as follows:

\begin{eqnarray}
\lefteqn{ ln\mathcal{L} = \sum_{i} ln P(L_{R_i}  | L_{X_i}, L_{K_i})\  -}  \nonumber \\
&& \sum_{j} \mathop{\int}_{L_R>L_{lim_j}}^{\infty} P(L_R  | L_{X_j}, L_{K_j}) dL_R.\nonumber\\
\end{eqnarray}

The first sum is computed for all the $i$ radio-detected sources and, as in eq. \ref{ML1},  is proportional to the combined probability of observing the radio luminosities L$_{R_i}$ of all the $i$ detected sources,
while the second term is the sum of the probability of radio detecting each $j$ observed (either radio detected or not) AGN, 
with a radio luminosity larger than its radio detection limit  $L_{lim_j}$. In this case, analogously to the classical maximum likelihood method (eq. \ref{ML1}), this second term corresponds to the expected  total number of radio detected sources.

	\subsection{The 3D Kolmogorov-Smirnov test }

Although the maximum likelihood technique is very powerful in finding the parameters of the best fit solution and their uncertainties, it doesn't allow to quantify how good the solution is.  We have then devised a three-dimensional  Kolmogorv-Smirnov (3D-KS)  test able to measure the probability of observing the 3D distribution (in the $L_R-L_X-L_K$ space) of our sample of 
radio detected sources, under the null hypothesis that the data are drawn  from the best fit model distribution.

The K-S test is a standard statistical test for deciding whether a set of data is consistent with a given probability distribution. In one 
dimension the K-S statistic is the maximum difference D between the cumulative distribution functions of the data
and the model (or another sample). What makes the K-S statistic useful is that its distribution (in the case of the null hypothesis that the data are drawn from the same distribution) can be calculated giving the significance of any nonzero value of D.

\begin{table*}

\begin{tabular}{c l c   c   c   c  c  c  c  c  c  c}

\hline\hline
Model & Function  & $\xi_X$ & $\xi_K$ & $\xi_0$ & Off-set & $\sigma_l^1$ & $\sigma_u$ & $a$ & $k$ & S   & P$_{KS}$ \\
\hline
1 & Gaussian               & 0.313 & 0.683 & 38.684 & ~0.00& 0.974 & ...      & ... & ... & 1583  &  2\% \\
2 & Double Gaussian & 0.373 & 0.582 & 38.513 & ~0.41 & 0.497 & 1.02 & ... &...  & 1571   &  3\% \\ 
3 & Lorentz                   & 0.379 & 0.705 & 39.374 & -0.43 & 1.141 & 0.30 &       ...   & ...        & 1515  &  26\%\\ 
\\
4 & Gaussian + exp    & 0.387 & 0.632 & 38.937 & ~0.09 & 0.582 &      ... & 0.117  & 1.66 & 1504 &  35\% \\
4 & 68\% conf. errors  & $^{+0.031}_{-0.059}$ & $^{+0.066}_{-0.053}$& $^{+0.050}_{-0.032}$ &  & $^{+0.028}_{-0.038}$ & & $^{+0.087}_{-0.021}$ & $^{+0.24}_{-0.07}$\\
\hline\hline
\multicolumn{6}{l}{1) Corresponding to $\sigma$ in those models with a single spread parameter.}\\
\end{tabular}
\caption{Best fit solutions}\label{tab:BestFitting}
\label{tabfit}
\end{table*}

In order to make a 3D-KS test we have followed the examples of the generalization of the K-S test to two-dimensional distributions by \citet{peacock83} and \citet{fasano87}. The maximum difference D statistic has been computed by measuring  the difference among the fraction of observed and expected (by the model) sources in each of the eight octants defined at the 3D positions of each radio detected source.
In order to compute the number of expected sources, 2000 Monte Carlo simulations have been used.
Although the 3D-KS test is carried out on the radio detected sample only,  it properly takes into accounts the effects of the radio
upper limits. Indeed, the simulations have been carried by extracting a radio luminosity (or not) for each observed source
by taking into account its radio detection limits an the model conditional probability distribution function $P(L_{R}  | L_{X}, L_{K})$. The same 2000 simulations have been used to compute the probability (significance) to observe the measured maximum D statistic in the case of the null hypothesis that the data are drawn from the same distribution.

	\subsection{The fits}
	
We have tried to fit the data assuming that the AGN radio luminosity, ${L_R}$, depends on average (i.e. with a spread in the radio luminosity axis) linearly from both the logarithmic X-ray and K-band (host galaxy) luminosities, drawing a plane in the 3D log$L_R$-log$L_X$-log$L_K$ space.
As already discussed in the introduction, this assumption is suggested by the observations, done by many authors, of the existence of a BH fundamental
plane: a similar relationship between the radio luminosity, the X-ray luminosity, and black hole mass ($M$), namely 
${\rm log} L_R = \xi_{RX} {\rm log} L_X + \xi_{RM} {\rm log} M + constant$ 
\citep[see][]{merl03,gult09}.
The AGN radio luminosity dependence on log$L_X$ and log$L_K$ 
 was modeled by the following relationship:
 
\begin{equation}
{\rm log}\bar{L_R} = \xi_X{\rm log}L_{X,44} + \xi_K{\rm log}L_{K,43} + \xi_0,
\label{eq_plane} 
\end{equation}

\noindent
where $\bar{L_r}$ is the luminosity of the peak ({\it mode}) of the probability distribution function of the spread in erg s$^{-1}$ units,
and the luminosities have been normalized to $L_X=10^{44} {\rm erg~s}^{-1} L_{X,44}$ and $L_K=10^{43} {\rm erg~s}^{-1} L_{K,43}$.

\begin{figure}
\begin{center}
\includegraphics[width = 8.5 cm]{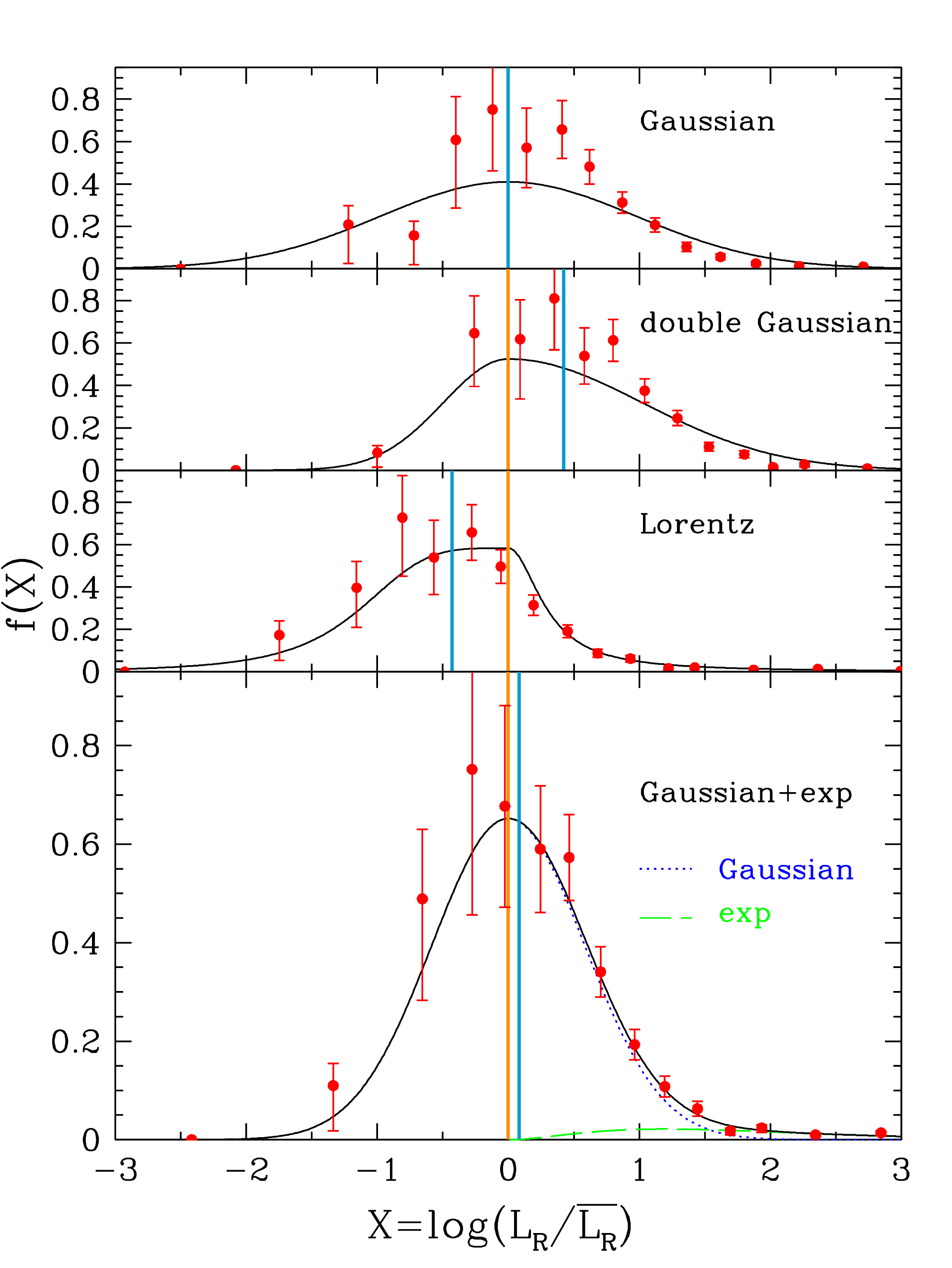}
\end{center}
\caption{ Probability distribution functions of $X={\rm log}(L_R/\bar{L_R})$, where $\bar{L_r}$ is the luminosity of the peak ({\it mode}), whose position is shown by a vertical continuous orange line.  The four distributions correspond to the four best fits reported in Table \ref{tabfit}. The blue vertical lines show the values of the means of the distributions, whose off-set from the
peak luminosity is reported in Table \ref{tabfit}, and which are used in the next figures to represent
the best fit solutions.}
\label{fig_models}
\end{figure}


This spread was first modeled by a 
Gaussian function, $G(X | 0, \sigma)$, where $X={\rm log}(L_R/\bar{L_R})$, centered on $X=0$ with standard deviation $\sigma$.   	
The best fit parameters are shown in Table \ref{tabfit} (model 1). 
However, the 3D-KS statistic tells that there is only a 2\% probability that the observed distribution is drawn from the model. 
The probability distribution function of $X$ is shown in Fig. \ref{fig_models}. 
The data have been plotted comparing in
each bin of $X$ the number of observed ($N_{obs}$) and expected ($N_{exp}$; by the model) sources. 
This method ($N_{obs}$ vs. $N_{exp}$ method) reproduces
the observations and consequently properly takes into
account both the radio detections and the upper limits \citep[see e.g.][for similar applications]{lafranca94, lafranca97,matute06}. 
It is worth to note that this fit, although not satisfactory, gives and indication that the PDF of L$_R$ is quite large: it results $\sigma\simeq 1$ dex. i.e. 68\% of the cases are included into a 2 dex wide distribution. Similar results have been found by 
\citet{merl03} and \citet{gult09}.

As shown in figure \ref{fig_models} \citep[and already observed by][]{lafranca10}, the data show an excess of high
radio luminosity sources if compared to a symmetrical distribution. In order to better take into account this excess we have
assumed an a-symmetrical double Gaussian distribution with two $\sigma$ values: $\sigma_l$ and $\sigma_u$ for radio luminosities values below (lower) and above  (upper) the value of the radio luminosity, $\bar{L_r}$, of the peak of the
probability distribution\footnote{defined by eq. \ref{eq_plane}.} (i.e. for $X$ values above or below zero) respectively. 
Indeed, the fit gives a larger spread, $\sigma_u$=1.0, at X$>$0,  than measured at X$<$0, where  $\sigma_l$= 0.5. 
However, even with this model, the 3D-KS test gives a not good enough, 3\%, probability.  
As the PDF is a-symmetrical, we show in Figure \ref{fig_models} both the locus of the peak and of the mean of the PDF. 
A better representation  of the model (e.g. in figures comparing the fit with the data) should, indeed, be carried out using the 
position of the  mean of the distributions. The offsets between the mean and the mode (off-set = mean - mode) of the PDF are also listed in column 4 of Table \ref{tabfit}.

Better results are obtained if the spread is modeled, as proposed by \citet{lafranca10}, assuming a double Lorentzian function described by the parameters $\sigma_l$ and $\sigma_u$:

\begin{figure}
\begin{center}
\end{center}
\includegraphics[width = 8.5 cm]{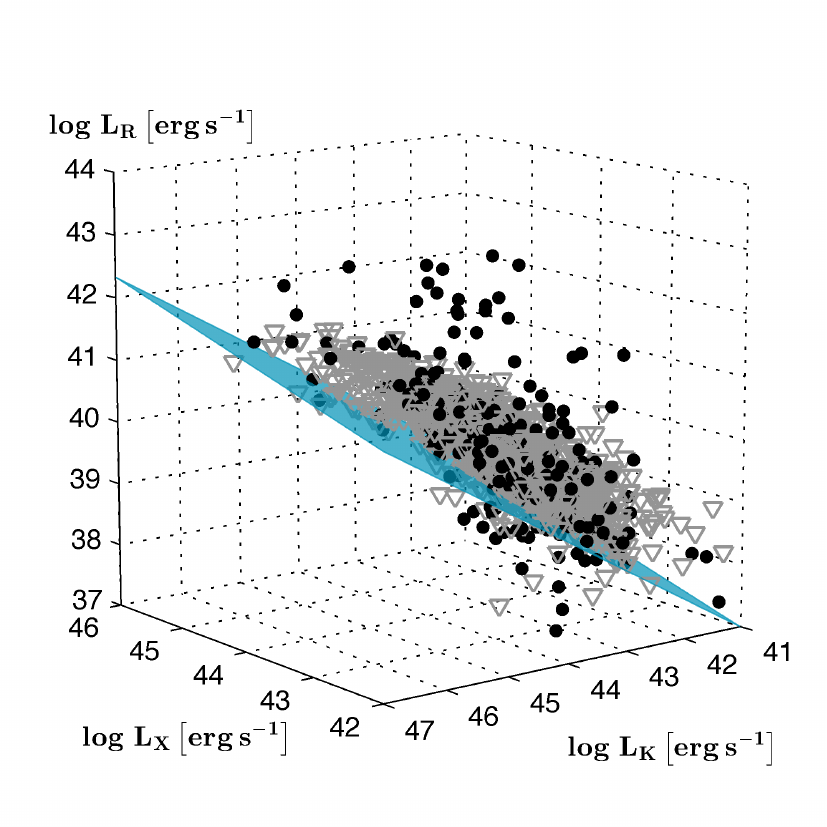}
\caption{3D distribution in the L$_R$-L$_X$-L$_K$ space. 
Filled circles represent radio detections and faint open triangles represent radio upper limits.
The plane of the distribution of the mean L$_R$ as a function of L$_X$ and L$_K$, according to the best fit solution 4 (Table \ref{tabfit}), is shown. }
\label{Fig3D}
\end{figure}

\begin{eqnarray}
P(X)=\left\{\begin{array}{l l}\frac{N}{A\pi\sigma_l \left[ 1 + \left(\frac{X}{\sigma_l} \right)^4 \right]}& \mbox{($X<0$)}\\ 
\\
\frac{A\ N}{\pi\sigma_u  \left[ 1 + \left(\frac{X}{\sigma_u} \right)^2 \right]}& \mbox{($X\geq 0$),}\\ 
\end{array} \right. \
\end{eqnarray}
\noindent
where,  in order to obtain a continuous function at $X$$=$$0$, $A = \sqrt{\sigma_u/\sigma_l}$, and
the parameter $N$ is constrained by the probability normalization requirement: $\int P(X)dX=1$.
In this case (see Figure \ref{fig_models} and model 3 in Table \ref{tabfit}) a 26\% 3D-KS probability is obtained.

An even better solution is obtained if,  for $X>0$, we add to  a Gaussian distribution, $G(X| 0,\sigma)$ (as in model 1),  an exponential function able to reproduce the high radio luminosity tail,

\begin{equation}
P(X)  = \left\{ \begin{array}{rl}
  bG(X|0,\sigma) & \mbox{ ($X<0$)} \\ 
 aX^2e^{-kX} + bG(X| 0,\sigma)  & \mbox{ ($X\geq0$),}  \\
       \end{array} \right.
\end{equation}
\noindent
 where the $a$ and $b$ parameters are not independent, as they are constrained by the probability
 normalization requirement

\begin{equation}
b = 1-\int_0^\infty  aX^2e^{-kX}dX.
\end{equation}

\begin{figure}
\begin{center}
\includegraphics[width = 6.5 cm, angle=-90]{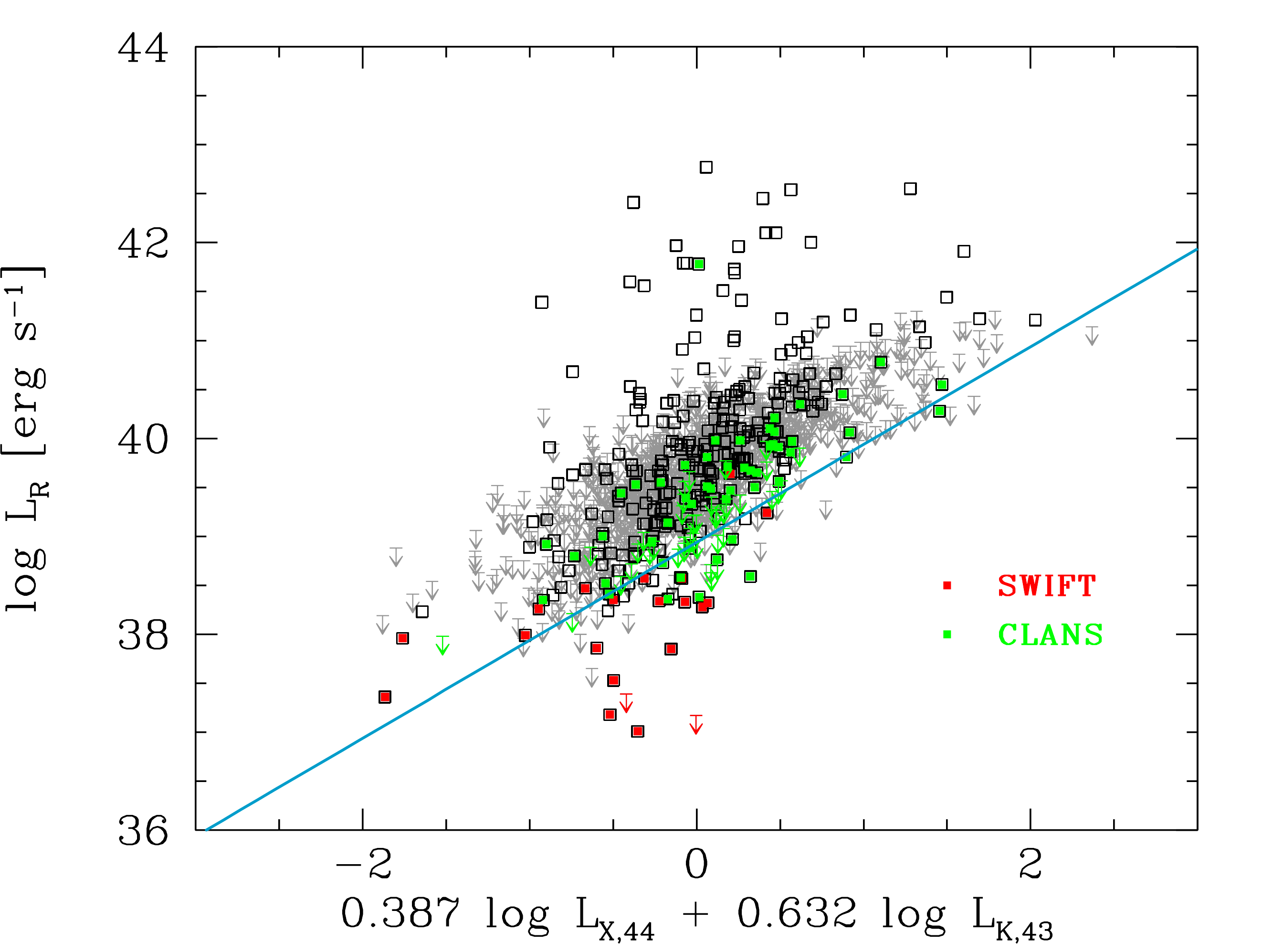}
\end{center}
\caption{Edge-on view of the 3D plane. The more complete samples, SWIFT and CLANS, are  represented by red and green squares, respectively. The continuous line shows the locus of the mean of the radio luminosity probability distribution function of our best fit solution 4 (Table \ref{tabfit}). Radio upper limits are represented by arrows.
}
\label{Fig_2D}
\end{figure}

\begin{figure}
\begin{center}
\includegraphics[width = 8.5 cm]{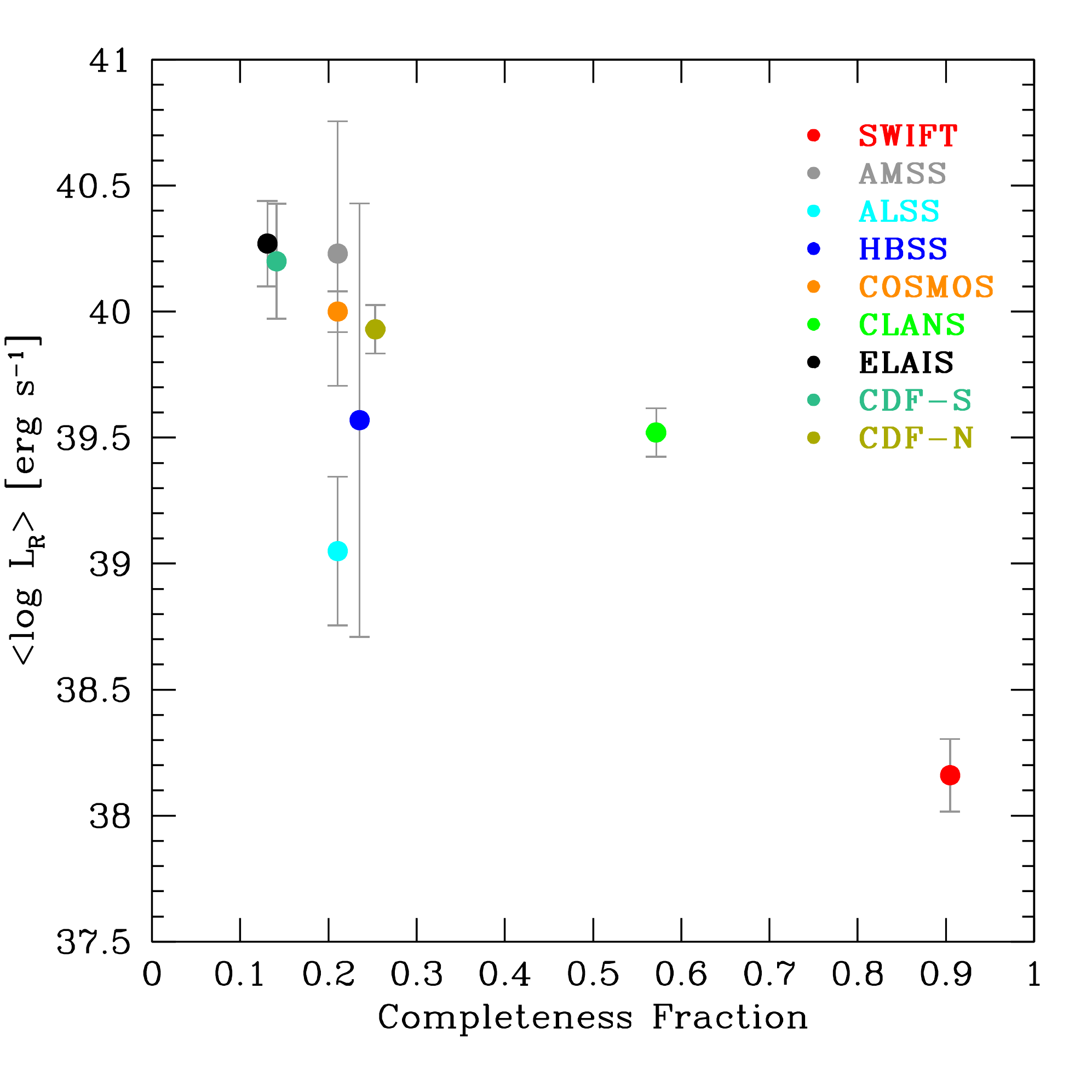}
\end{center}
\caption{Mean radio luminosity of the radio detected sources of each sample as a function of the radio detection completeness. 
}
\label{fig_RadioMean}
\end{figure}

The best fit solution  (model 4 in Table \ref{tabfit} and see Figure \ref{fig_models}) gives b=0.9499, which implies that the added exponential tail at high radio luminosities represents
about 5\% (1-0.9499) of the population (10\% for X$\geq$0 where it is defined). In this case a 35\% 3D-KS probability is obtained.
As already shown by \citet{lafranca10} the radio luminosity distribution of  the AGN does not show any evidence of a bimodal
distribution and therefore any definition of radio loudness is arbitrary.
Nonetheless, it should be observed that (as demonstrated by these fits) the radio luminosity PDF is not symmetrical, but skewed with a long tail
at high radio luminosities. The introduction of this tail at X$>$0 allows to  obtain a narrower ($\sigma\sim 0.6$ dex) complementary symmetrical Gaussian distribution.
The position of the peak of the radio distribution of our best fit solution (number 4) is represented by the following equation:

\begin{equation}
{\rm log}\bar{L_R} = 0.39^{+.03}_{-.06}{\rm log}L_{X,44} + 0.63^{+.07}_{-.05}{\rm log}L_{K,43} + 39.94^{+.05}_{-.03},
\label{eq_fit} 
\end{equation}

\begin{figure}
\begin{center}
\end{center}
\includegraphics[width = 8.5 cm]{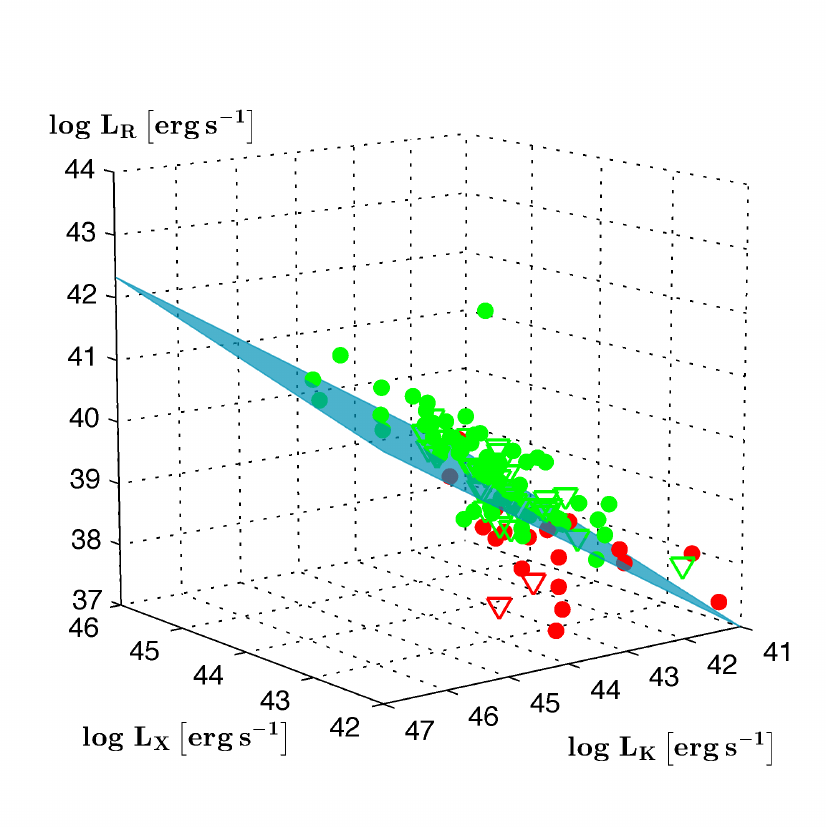}
\caption{Same as Figure \ref{Fig3D} but using only the, more radio complete, SWIFT (red circles) and CLANS (green circles) samples. Radio upper limits are represented
by open triangles.
The plane of the distribution of the mean L$_R$ as a function of L$_X$ and L$_K$, according to the best fit solution 4, is shown.}
\label{Fig3Dsub}
\end{figure}

 Confidence regions of each
parameter were obtained by minimizing the $S$ function at a number of values around the best-fit solution, while leaving the other parameters free to float \citep[see][]{lampton76}. The 68\% confidence regions quoted correspond to $\Delta S$ (=$ \Delta \chi^2$) = 1. 
The best fit solution is also shown in Figure \ref{Fig3D},
where the 3D distribution of the data is shown, and Figure \ref{Fig_2D},  where the 2D edge on view of the plane is shown.
This last figure helps understanding how much important is to take into account the effects of using censored data. 
The fitting solution seems, indeed, to be not a good representation of the distribution of the radio
detections. This is because most of the X-ray samples have no radio observations able to detect all sources
and then the radio detections are biased in favor of the most luminous radio sources. Indeed the 
mean radio luminosity of each of our samples decreases as a function of the radio identification completeness fraction: the average radio luminosity of the samples with about 10\% radio identifications is of about logL$_R$=40.0 erg/s, while for the
90\% complete samples the average luminosity is about 2 dex lower  (logL$_R$=38.2 erg s$^{-1}$; see Figure \ref{fig_RadioMean}). This bias is properly taken into account with our ML fitting method which takes into account upper limits. The fitting solution, indeed, fairly well reproduces the distribution of the most radio complete samples, such as SWIFT and CLANS (see Figures \ref{Fig_2D} and \ref{Fig3Dsub}).

\section{The AGN fundamental plane}

Many authors have observed the existence of a BH fundamental
plane relationship between the radio luminosity, the X-ray luminosity, and the BH  mass (see the discussion in the Introduction). As  a relation has been observed between the BH mass and the bulge luminosity, 
it is interesting to study whether  the BH fundamental plane measures are compatible or not with
our measured relationship in the log$L_R$-log$L_X$-log$L_K$ plane (we remember that L$_K$ is the galaxy, stellar component, luminosity).
We have therefore estimated the BH masses using the calibrated black hole versus K-band bulge luminosity relation from \citet{graham07}:
\begin{equation}
log(M_{bh}/M_{\odot}) = -0.37(M_{K, bulge}+24) + 8.29.
\label{eq_Gr07}
\end{equation}
The bulge to total luminosity ratio (B/T), or the bulge to disc ratio (B/D; where $B/T=[1+D/B]^{-1}$) are function of the galaxy type \citep[see][]{dong06,graham08}.  Therefore,
in order to properly derive the bulge luminosities from our measures of the total galaxy luminosities it would be necessary to  know the morphological type of our galaxies.  Unfortunately this information is barely available only for few tenths of galaxies belonging to the local sample (SWIFT), while for the higher redshift galaxies no information is available.
Moreover, such a kind of relationships have been calibrated only on local samples, while it is well known that the average
galaxy mass and morphology changes with redshifts \citep[e.g. smaller and more clumpy galaxies with increasing redshift;][]{mosleh12}, and 
our sample reaches $z\sim5$.  However, our aim is not to measure the BH fundamental plane, but only to verify
how much compatible our measure is with previous measures.  According to  \citet{graham08}, we have therefore assumed an average value of $B/T=1/4$ in the K bandpass \citep[see e.g.][for similar assumptions]{fiore12}. Under this assumption, eq. \ref{eq_Gr07} corresponds to the relation

\begin{equation}
{\rm log}(M_{bh}/M_{\odot}) = 0.925 {\rm log} L_{K} -31.781,
\label{Eq_mbh}
\end{equation}

\noindent
where L$_{K}$ is the total (bulge plus disk) galaxy luminosity, expressed in erg s$^{-1}$ units. We have compared 
our BH mass estimates with those reported by \citet{merloni10}, which have been obtained via virial based analysis of optical spectra of AGN1. 
On average, our estimates are larger by 0.14 dex in solar masses with a spread of 0.43 dex, in solar mass units. This 
spread is compatible with the typical spreads ($\sim$0.3-0.4 dex) of the BH mass estimates based on both the virial and the scaling relations methods \citep{gultekin09},  whose  uncertainties, in our comparison, should be both taken into account in the propagation of errors.

According to eq. \ref{Eq_mbh}, our best fit solution 4 is transformed into the L$_R$-L$_X$-M$_{BH}$ space  in to the relation

\begin{equation}
{\rm log} L_R = 0.39 {\rm log} L_X + 0.68 {\rm log} M_{BH} + 16.61
\end{equation}

\noindent
while \citet{merl03} measure  

\begin{equation}
{\rm log} L_R = 0.60 {\rm log} L_X + 0.78 {\rm log} M_{BH} + 7.33,
\end{equation}

\noindent
where L$_R$ here is the 5 GHz nuclear luminosity in units erg s$^{-1}$,
with 1.4 GHz radio luminosities converted into 5 GHz $\nu$L$_\nu$ luminosities assuming a radio spectral index $\alpha=0.7$ \citep[where L$_\nu \propto \nu^{-\alpha}$;][]{condon02}, L$_X$ the 2-10 keV nuclear X-ray luminosity in units of erg s$^{-1}$, and M$_{BH}$ the black hole's mass in units M$_\odot$ \citep{merl03}.

\section{Discussion and Conclusions}

\begin{figure}
\begin{center}
\includegraphics[width = 8.5 cm]{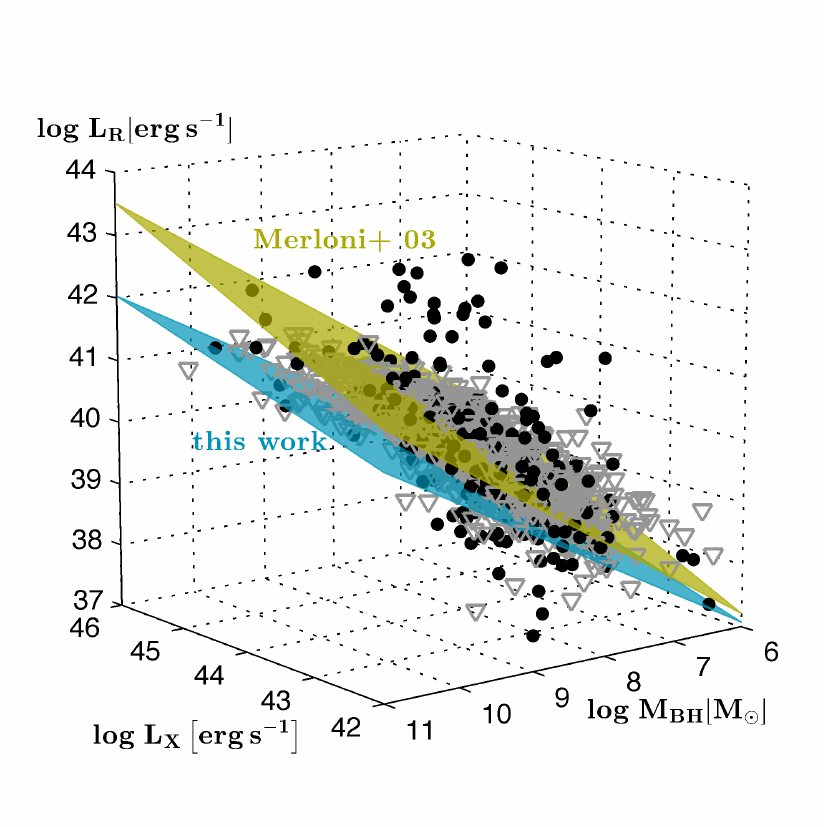}
\end{center}
\caption{
3D distribution in the in L$_R$-L$_X$-M$_{BH}$ space. The  projection from the L$_R$-L$_X$-L$_{K}$ space
of the plane of the best fit solution 4 is shown in cyan.  The fundamental plane from \citet{merl03}  is shown in gold.
Radio upper limits are represented by open triangles.}
\label{Fig_Fund}
\end{figure}

\begin{figure}
\begin{center}
\includegraphics[width = 8.5 cm]{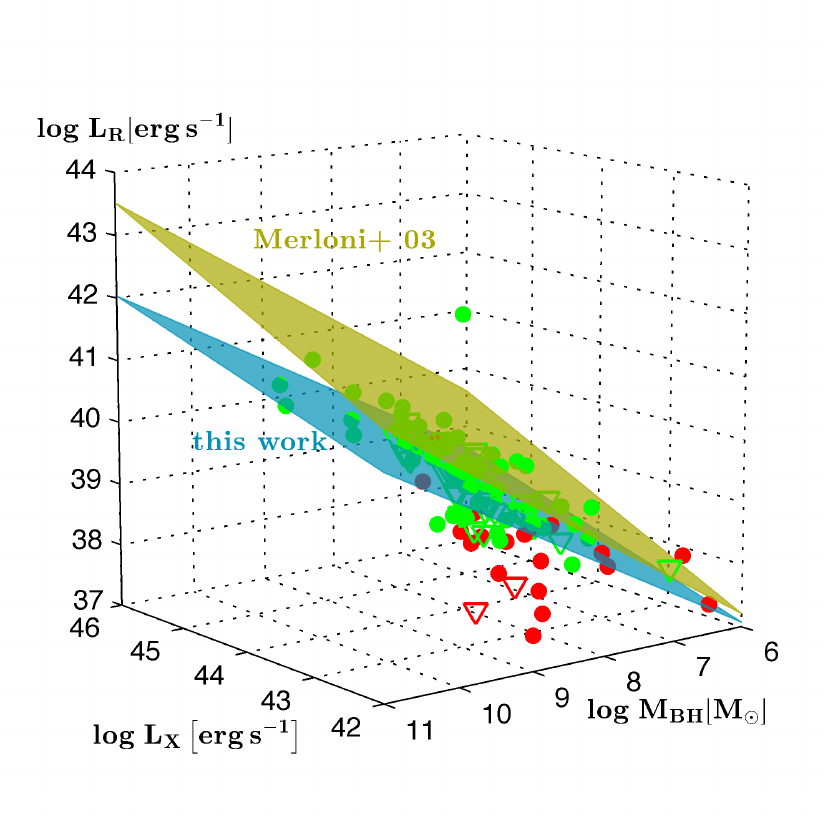}
\end{center}
\caption{Same as Figure \ref{Fig_Fund} but 
using only the, more radio complete,  SWIFT (red circles) and CLANS (green circles) samples. Radio upper limits are represented by open triangles.}
\label{Fig_FundSub}
\end{figure}

\begin{figure}
\begin{center}
\includegraphics[width = 6.5 cm, angle=-90]{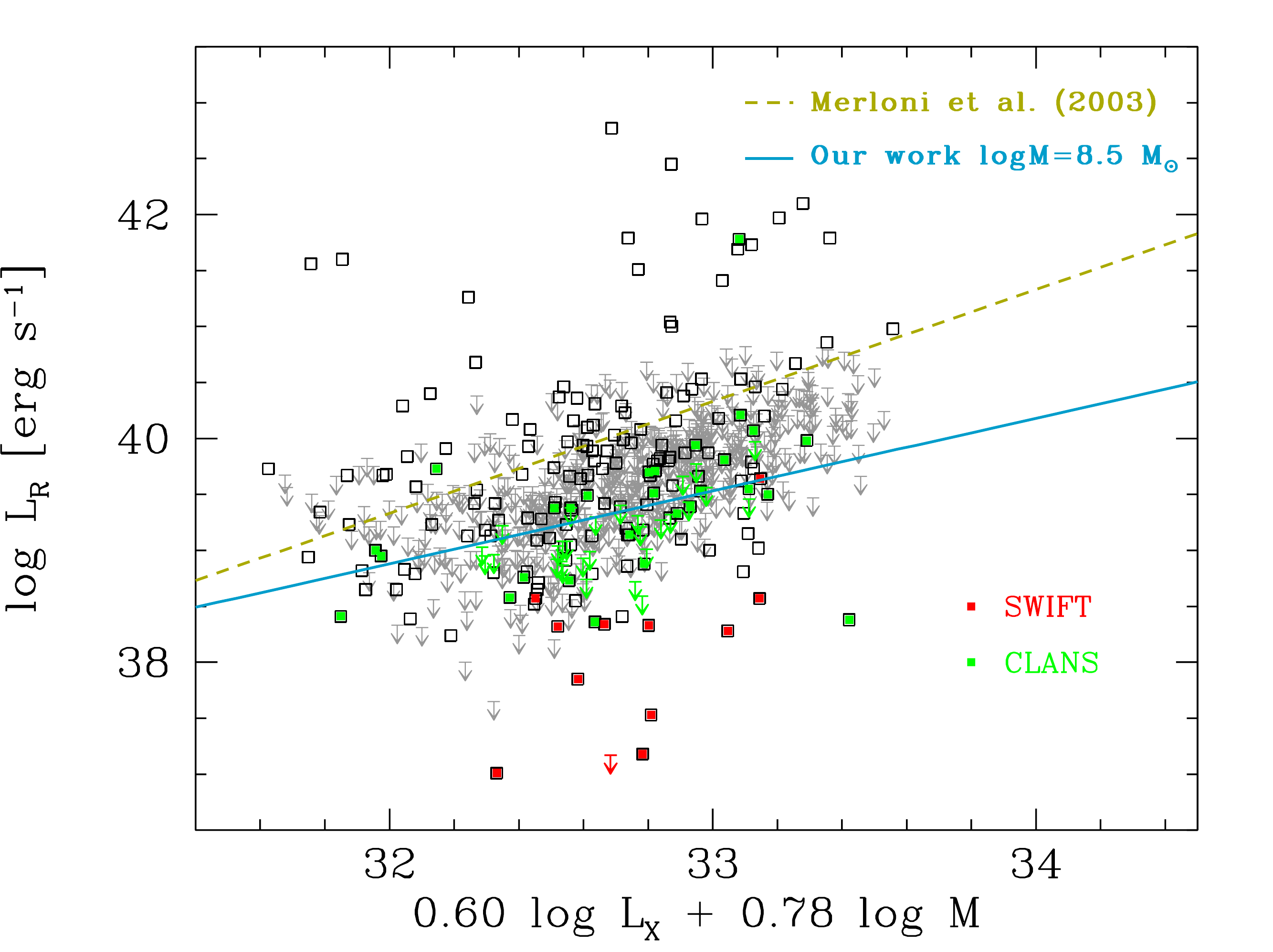}
\end{center}
\caption{Edge on view of the BH fundamental plane as measured by \citet{merl03} (gold dashed line). The continuous cyan line
shows our solution for a sample of AGN having average logM=8.5.
The more complete samples, SWIFT and CLANS, are  represented by red and green squares, respectively. Radio upper limits are represented by arrows.}
\label{Fig_2DRXM}
\end{figure}

As expected (see Figure \ref{Fig_Fund}), our estimate of the BH fundamental plane predicts lower radio luminosities if compared with the previous measure by \citet{merl03}. The typical difference, computed at 10$^8$ M$_\odot$ BH mass and 10$^{44}$ erg s$^{-1}$ X-ray luminosity, is of $\sim$0.8 dex. 
As already observed in the L$_R$-L$_X$-L$_{K}$ space, this difference is due to the inclusion in our analysis of the contribution of the radio upper limits, and indeed  our best fit solution  reproduces well the distribution of the most radio complete samples such as SWIFT and CLANS (Figures \ref{Fig_FundSub} and \ref{Fig_2DRXM}).

It should be noted that the fundamental plane of \citet{merl03} was constructed by including a sample of X-ray BH binaries. While it is interesting to see that the mass scaling is still broadly consistent with that of \citet{merl03} even in our study of an AGN-only sample, the radio-X-ray coefficient is very different: $\sim$0.4 in this study, while low-accretion rate (and low-luminosity) X-ray BH binaries show a tight radio-X-ray correlation with slope $\sim$0.6. Interestingly enough, recent re-analysis of radio-X-rays correlations in X-ray binaries suggests the presence of a second, less radio luminous branch 
\citep[][]{gallo03,coriat11,gallo12}. In this framework, our study, could suggest that, also in AGN, a second less radio luminous population
should be taken into account, which could corresponds to those objects with low values of the  $X={\rm log}(L_R/\bar{L_R})$ parameter (see Figure \ref{fig_models}).

As already discussed in the Introduction, the measure of the dependence in AGN of
L$_R$ from other physical quantities such as L$_X$ and the host galaxy K-band
luminosity, L$_K$, is very useful in order to better include AGN in the
galaxy evolution models where the  AGN/galaxy feedback play a relevant role.

Our analysis has allowed to find a good analitical solution represented by a plane in the
3D logL$_R$-logL$_X$-logL$_K$ space once a wide (1$\sigma$$\sim$1 dex), a-symmetrical, spread in the
radio luminosity axis is included.
This result confirms the study of \citet{lafranca10}, who, studying the dependence of PDF of L$_R$ from
L$_X$ and $z$, where able to model the 1 dex wide (1$\sigma$) spread in the AGN radio luminosity distribution.
This results show that a proper study of the correlation between different bands luminosities in AGN (or other sources) cannot be performed without taking into account censored data. 

A clear example, in this framework, is the measure of the BH fundamental plane in the 3D logL$_R$-logL$_X$-logM space.
After converting, using scaling relations, our measures of the host galaxy K-band luminosity into BH masses, our
best fit solution corresponds to a BH fundamental plane which on average predicts 0.8 dex lower values for the AGN
radio luminosities.

It should be pointed out that, at variance with many similar statistical studies, our analysis is  based on a compilation of complete, hard X-ray selected,  AGN samples, where both AGN1 and AGN2 are included. Therefore, our results should better
represent the behaviour of the whole AGN population.
However
it will be interesting to check the 3D correlation between L$_R$, L$_X$ and L$_K$ in complete, {\em radio selected} samples. As the \citet{merl03} sample was a hybrid sample without a clear selection criterion,  it is plausible that parts of the differences found in the present work could be due to the specific selection criterion. In very general terms, if we do not believe we know any of the three terms (X-ray luminosity, radio luminosity and BH mass) as being the primary physical driver, all should be treated equal in a correlation study (but this is beyond the purposes of this paper).

In order to improve these analysis it would be very useful to obtain deeper radio observations of complete
samples of AGNs combined with detailed optical-NIR-MIR SED observations.
These data, when available, will eventually allow to measure the dependence of the radio luminosity distribution (i.e. the feedback) from the star and BH masses, their derivatives (star formation and accretion rates) and the redshift. A result
that could be achieved by complementing  multiwavelength surveys with observations carried out with new radio facilities  such as the Expanded Very Large Array and the Square Kilometer Array.

\section{Acknowledgments}

We thank Andrea Merloni and Geoffrey Bicknell for discussions.
We acknowledge the referee for his very careful review that allowed us to improve the quality of this work.
This publication  uses the NVSS and the FIRST radio surveys, carried out using the National Radio Astronomy
Observatory Very Large Array. NRAO is operated by Associated University Inc., under cooperative agreement with the National Science Foundation.
This publication makes use of data products from the Two Micron All Sky Survey, which is a joint project of the University of Massachusetts and the Infrared Processing and Analysis Center/California Institute of Technology, funded by the National Aeronautics and Space Administration and the National Science Foundation. We acknowledge financial contribution from PRIN-INAF  2011.

\bibliographystyle{mn2e}
\bibliography{sbs}

\label{lastpage}

\end{document}